\documentclass{article}

\usepackage[utf8]{inputenc} 
\usepackage[T1]{fontenc}    
\usepackage{hyperref}       
\usepackage{url}            
\usepackage{booktabs}       
\usepackage{amsfonts}       
\usepackage{nicefrac}       
\usepackage{microtype}      
\usepackage{graphicx}       

\usepackage[left=1.2in, right=1.2in, top=1in, bottom=1in]{geometry}
\usepackage[style=nature]{biblatex} 
\usepackage[labelfont=bf,textfont=md]{caption}
\usepackage{algorithm, algpseudocode}
\usepackage{authblk}
\usepackage{chemformula}

\usepackage[draft]{changes}

\title{Spectroscopy-Guided Discovery of Three-Dimensional Structures of Disordered Materials with Diffusion Models}

\author[1]{Hyuna Kwon\thanks{kwon11@llnl.gov, equal contribution}}
\author[1]{Tim Hsu\thanks{hsu16@llnl.gov, equal contribution}}
\author[1]{Wenyu Sun}
\author[1]{Wonseok Jeong}
\author[1]{Fikret Aydin}
\author[2]{James Chapman}
\author[1]{Xiao Chen}
\author[3]{Matthew R. Carbone}
\author[4]{Deyu Lu}
\author[1]{Fei Zhou}
\author[1,5]{Tuan Anh Pham\thanks{pham16@llnl.gov}}
\affil[1]{Lawrence Livermore National Laboratory, Livermore, CA 94551, USA}
\affil[2]{Department of Mechanical Engineering, Boston University, Boston, MA 02215, USA}
\affil[3]{Computational Science Initiative, Brookhaven National Laboratory, Upton, NY 11973, USA}
\affil[4]{Center for Functional Nanomaterials, Brookhaven National Laboratory, Upton, NY 11973, USA}
\affil[5]{Laboratory for Energy Applications for the Future, Lawrence Livermore National Laboratory, Livermore, CA 94551, USA}
\date{}
\begin{document}
\maketitle
\begin{abstract}
The ability to rapidly develop materials with desired properties has a transformative impact on a broad range of emerging technologies. In this work, we introduce a new framework based on the diffusion model, a recent generative machine learning method to predict 3D structures of disordered materials from a target property. For demonstration, we apply the model to identify the atomic structures of amorphous carbons ($a$-C) as a representative material system from the target X-ray absorption near edge structure (XANES) spectra---a common experimental technique to probe atomic structures of materials. We show that conditional generation guided by XANES spectra reproduces key features of the target structures. Furthermore, we show that our model can steer the generative process to tailor atomic arrangements for a specific XANES spectrum. Finally, our generative model exhibits a remarkable scale-agnostic property, thereby enabling generation of realistic, large-scale structures through learning from a small-scale dataset (i.e., with small unit cells). Our work represents a significant stride in bridging the gap between materials characterization and atomic structure determination; in addition, it can be leveraged for materials discovery in exploring various material properties as targeted.

\end{abstract}

\section{Introduction}

An accelerated materials discovery pipeline with desired properties holds a transformative impact on a broad range of emerging technologies, from energy storage and conversion to ion-selective membranes~\cite{simon2008materials,nitta2015li,aluru2023}. Traditional approaches for materials discovery involve a time-consuming trial-and-error process that begins with a set of materials from which properties of interest are derived. The understanding gained from this procedure is then used to guide materials development by exploring a new configuration space. Recent advancement in data science approaches provides an alternative and promising means to accelerate materials development~\cite{correa2018accelerating}. For example, high-throughput materials screening and data mining have led to several breakthroughs in materials development for a broad range of technologies in the past two decades~\cite{jain2011high,jain2013commentary}. This approach often relies on simulations techniques, such as first-principles calculations, to rapidly screen a vast space of chemical compositions and structures in order to identify materials with desired properties. 

Materials development through inverse design departs from the traditional methods. Instead of deriving properties from structures, target material properties are specified beforehand and materials satisfying these requirements are inferred~\cite{sanchez2018inverse}. This strategy, however, is generally more challenging. For example, many material systems exhibit multi-to-one mapping in the structure-property space, i.e., multiple distinct structures can result in roughly the same property. It is therefore challenging to learn the inverse mapping even if the forward mapping is fully known. This problem has been the topic of active research in recent years with data-driven approaches, thanks to emerging capabilities of modern machine learning methods~\cite{noh2019inverse,yao2021inverse,kim2020inverse,noh2020machine}. A viable direction is to formulate the inverse prediction problem in a probabilistic framework, for instance with a conditional generative model, that generates or samples multiple candidate material structures conditioned on a given property. 

Among the different classes of generative models, the diffusion model \cite{Sohl-Dickstein2015-DPM, Ho2020-DDPM, Song2020-unified} is particularly well suited for generating structures involving precise three-dimensional (3D) atomic coordinates. The diffusion formulation centers around the idea of denoising perturbed or noisy prior inputs towards realistic distributions based on a learned score function \cite{Hyvarinen2005, Vincent2011scorematching} that has the form of a relative distance vector for atomic coordinate data. In addition, when the data space is defined in terms of relative distance vectors between atoms, the score model does not rely on absolute coordinates and is therefore translation-invariant. While there have been numerous works on diffusion model for molecules \cite{hoogeboom2022equivariant, jing2022torsional, vignac2022digress, xu2022geodiff, weiss2023guided}, relatively few studies have focused on disordered systems~\cite{xie2021crystal, zheng2023towards}. This family of materials is highly relevant in many emerging applications, such as energy conversion and storage devices, where the materials are often driven away from ideal structures under external driving forces~\cite{pham2017modelling,wood2020beyond}. They can therefore be more complex, often exhibiting richer structural and chemical heterogeneities.

In this study, we demonstrate the application of the diffusion model to identify 3D structures of disordered solid materials from a target property, which in this work is chosen to be spectroscopy to address an outstanding challenge in materials characterization: predicting 3D structures of disordered materials from spectroscopic measurements. We focus on XANES spectroscopy that is widely used to probe local atomic structures, and demonstrate with amorphous carbons ($a$-C) as a representative system. In a similar vein, Comin and Lewis \cite{comin2019deep} trained a Wasserstein autoencoder~\cite{tolstikhin2017wasserstein} to generate amorphous silicon topologies. However, such a model is limited to a fixed number of atoms whereas our diffusion model implemented with graph neural networks (GNN) is scale-agnostic, allowing for generation at any arbitrary scale. We stress that although our present study focuses on single-element $a$-C and XANES spectroscopy, the conditional diffusion model framework can be extended to broader scenarios. For instance, the scale-agnostic property of the diffusion model allows for the generation of large-scale atomic structures based on information learned from a set of diverse small-scale data. The framework can also be naturally extended to multi-element material systems, where the atom type information is also subject to denoising. Finally, multi-target inverse prediction is also straightforward to implement to enable material inverse design for various target properties, as the conditional score function is a simple summation of the prior and likelihood scores from multiple forward models.

\begin{figure}
    \centering
    \includegraphics[width=0.8\textwidth]{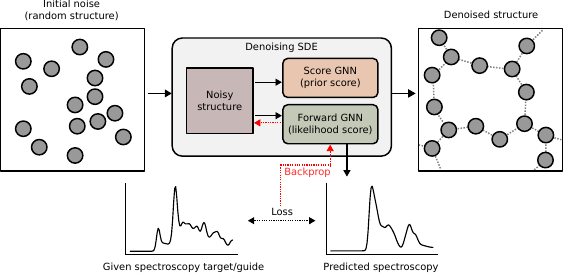}
    \caption{The conditional generative/denoising process of amorphous carbon ($a$-C) involves a pair of decoupled (i.e., trained separately) score model and forward model. The score model, based on the diffusion model formalism, provides the prior score, while the forward model, trained to predict spectroscopic curves, is used to provide the likelihood score. The prior score serves to denoise initial random motifs towards realistic topologies without bias, and the likelihood score serves to bias the denoising trajectory towards atomic motifs that satisfy a given spectroscopic target/guide. Notably, the forward model is optional, and the generative process is simply unconditional without a forward model.
    }
    \label{fig:intro}
\end{figure}

Our conditional diffusion model framework shown in Fig.~\ref{fig:intro} follows a standard structure consisting of a prior score model and a forward model. The former outputs the \textit{prior/unconditional score}, which is used to gradually denoise noisy priors towards realistic samples without bias, while the latter---which predicts per-atom spectral curves in this work---provides the \textit{likelihood/conditioning score} that guides the denoising iterations towards a given target property. It is necessary to emphasize that the target property is not necessarily always achievable (and not all targets are realistic). In this regard, the given spectroscopy target may be more appropriately described as a guide. Importantly, the prior score model and the forward model are decoupled and can be separately trained. We note that the use of a forward model is optional, in which case the generation is simply unconditional. Further details regarding the conditional diffusion model implementation are described in the Methods section.

\section{Results}

We present our results regarding $a$-C 3D structure generation with and without XANES spectroscopy conditioning. In the unconditional case, we focus on the fidelity of the generation, and highlight that the generation is scale-agnostic. In the conditional case, the performance of a structure-to-spectroscopy forward model is briefly discussed, and the spectroscopy-guided generation is validate in two scenarios. The extent of the conditioning effect is also briefly analyzed. Overall, we demonstrate realistic, scale-agnostic generation of $a$-C structures based on spectra inputs/guides.

\subsection{Unconditional generation}

The fidelity of the structure generation is first examined without considering the conditioning effect. We emphasize that our generation model is scale-agnostic, i.e., the model can learn from small-scale molecular dynamics (MD) trajectories and consequently generate similar configurations at any arbitrary scale. This is enabled by the node-centric (local) predictions of the underlying graph neural network architecture. The denoising process therefore amounts to shifting each atom's coordinate towards realistic topologies based on its local environment without the need to observe global variables. This scale-agnostic feature is particularly useful for generating large domains encompassing diverse topologies, which can be costly to sample via MD simulations. In this regard, the large-scale generated structures can be interpreted as having ``stitched'' or combined the small but diverse MD-simulated motifs. Such a generation is important for examining material properties where a large supercell with diverse local structures is required for yielding sufficient statistics. Examples include predicting hydrogen diffusion~\cite{chapman2023hydrogen}, gas selectivity~\cite{jiang2013molecular}, interfacial interactions \cite{gaikwad2022molecular}, and graphitization~\cite{li2018reaxff} in amorphous systems and disordered interfaces. We intend to investigate scaled generations as part of future effort. 

\begin{figure}
    \centering
    \includegraphics[width=\textwidth]{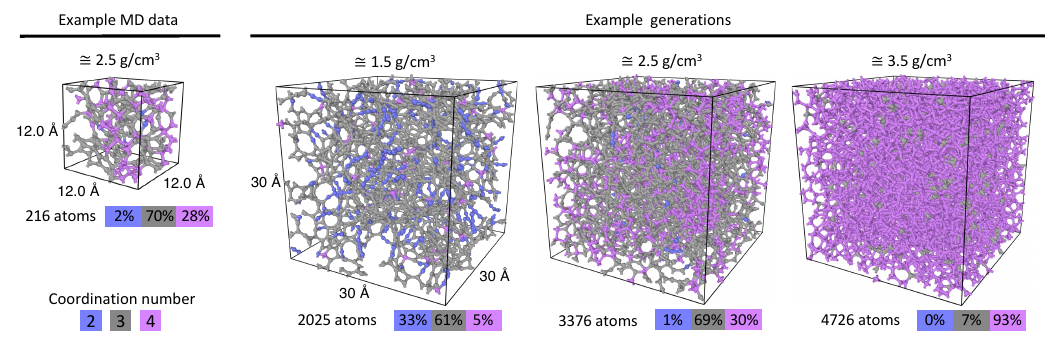}
    \caption{The generation of the $a$-C structures is scale-agnostic, and the generated topologies highly depend on  density (as expected). While the leftmost panel displays a representative training data point with a density of 2.5~g/cm$^3$, it is essential to note that the subsequent larger cells on the right are not exclusive products of this single 2.5~g/cm$^3$ cell. The generation can be at any arbitrary scale because the denoising process amounts to shifting each atom's location towards realistic topology based on its local environment. Consistent with the MD training data, the $a$-C topologies generated at low density typically have a small coordination number (CN) of 2 and 3, and at high density the CN is typically 3 and 4.   
    }
    \label{fig:gen-visual}
\end{figure}

The scale-agnostic feature is demonstrated in Fig.~\ref{fig:gen-visual}, where the generated $a$-C configurations have a supercell 2.5 times larger than that of the original training data. As expected, the generated $a$-C topology highly depends on the density (or the number of atoms per unit volume) at which the structure is generated. Specifically, $a$-C motif tends to be $sp$- or $sp^2$-hybridized at low densities, and $sp^2$- or $sp^3$-hybridized at high densities. In fact, the generation is always (implicitly) conditioned on the density (input cell size and number of particles within) even without a spectral target. For simplicity, the term \textit{unconditional} used throughout this paper refers to generation without a spectral target. We note that the coordination number (CN) is used to estimate the hybridization. Specifically, we define two atoms as bonded if their distance is less than a cutoff of 1.84~Å, which is determined based on the C-C radial distribution function (RDF) of the structures in the training set.

\begin{figure}
    \centering
    \includegraphics[width=0.8\textwidth]{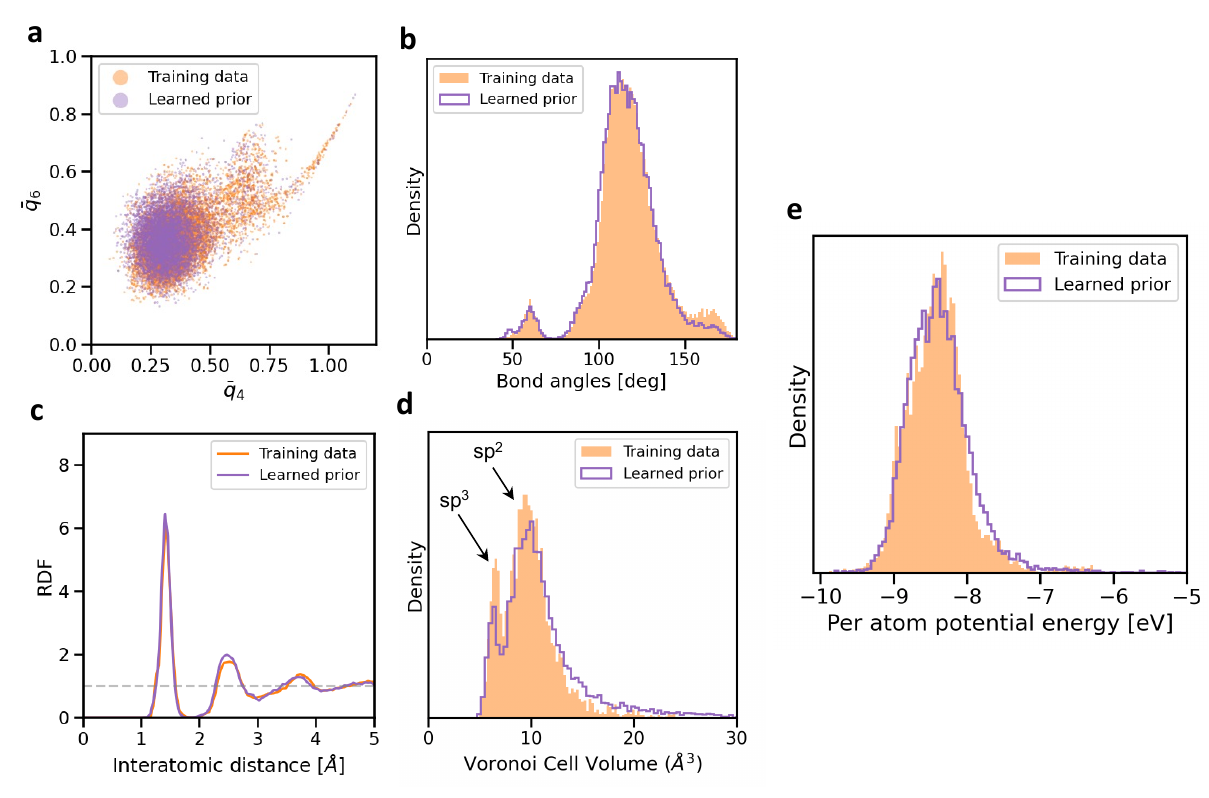}
    \caption{The generated prior amorphous topologies are realistic based on comparisons of (a) Steinhardt parameters \cite{steinhardt1983bond}, (b) bond angle distribution,  (c) radial distribution, (d) Voronoi cell volume distribution, and (e) per-atom GAP potential energy distribution \cite{caro2020GAP, deringer2017machine}  against the training data.
    The generated structures consist of 29 cubic cells ($14^3$Å) of $a$-C, totaling 7,256 atoms.}
    \label{fig:prior-fidelity}
\end{figure}

The unconditionally generated structures are realistic (Fig.~\ref{fig:prior-fidelity}), adhering to the training data distributions in terms of the various structural and thermodynamic properties considered in this work: Steinhardt order parameters \cite{steinhardt1983bond}, bond angles, RDFs, Voronoi cell volumes, and per-atom  potential energies \cite{caro2020GAP, deringer2017machine}. The Steinhardt parameter in particular stands out for its detailed insights into the local symmetry of atomic arrangements, complementing other descriptors like RDF and bond angle distribution. First, as shown in Fig.~\ref{fig:prior-fidelity}a, the scatter plot of the Steinhardt features $(\bar{q}_4, \bar{q}_6)$ indicates that the scatter points from the generated $a$-C structures roughly match the distribution of the training data, but importantly they do not coincide. This indicates that the generated topologies are similar to, but not exact copies of, the training atom motifs. Second, the bond angle, RDF, and Voronoi cell volume distributions further validate that the generated configurations are realistic, resembling the configurations of the training set (Fig.~\ref{fig:prior-fidelity}b-d). Finally, Fig.~\ref{fig:prior-fidelity}e shows that the per-atom potential energy distribution obtained from the carbon Gaussian Approximation Potential (GAP)~\cite{deringer2017machine} of the generation largely matches that of the training data. This observation suggests that the generated structures are not only geometrically realistic, but also energetically realistic. Importantly, the training dataset consists of 30 snapshots (of various cell sizes) of MD-simulated $a$-C, totaling 7,463 carbons. Overall, the $a$-C generation is realistic, exhibiting quantities highly similar, but not identical, to that of the MD data. It is viable to subject the generation to downstream atomistic simulations thanks to the generated atoms having realistic energies.

\subsection{Conditional generation}

\begin{figure}
    \centering
    \includegraphics[width=0.8\textwidth]{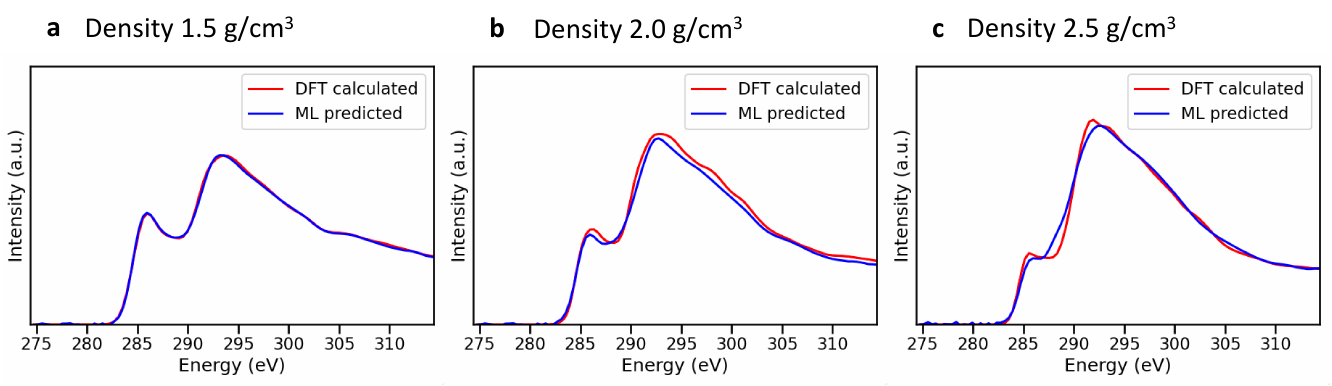}
    \caption{The forward XANES spectroscopy regression model is sufficiently accurate on unseen validation $a$-C configurations. Note that the forward model predicts per-atom spectral curves, which are then averaged to give rise to global spectra that are compared to DFT calculations. The comparison is shown  for a validation set of three structures at densities (a) 1.5, (b) 2.0, and (c) 2.5 g/cm³.
    }
    \label{fig:forward_model}
\end{figure}

Since the conditional generation formulation requires a forward prediction model, we trained a forward model to predict XANES spectra of $a$-C configurations from their atomic structures (Supplemental Fig.~\ref{fig:forward_model_scheme}). Illustrated in Fig.~\ref{fig:forward_model}, this model is sufficiently accurate on three validation $a$-C structures that yield distinctively different XANES spectra (at a density of 1.5, 2.5, and 3.5 g/cm$^{3}$, respectively). Both the forward model and the unconditional score model are based on the same GNN architecture, and further details regarding the forward model are described in the Methods section. We point out that in our prior work~\cite{kwon2023harnessing} we trained a forward model based on a multi-layer perceptron (MLP) architecture with Many Body Tensor Representation (MBTR) descriptors as the input, and the current GNN exhibits a marginal yet discernible performance improvement over the MLP-based model.

\begin{figure}
    \centering
    \includegraphics[width=0.9\textwidth]{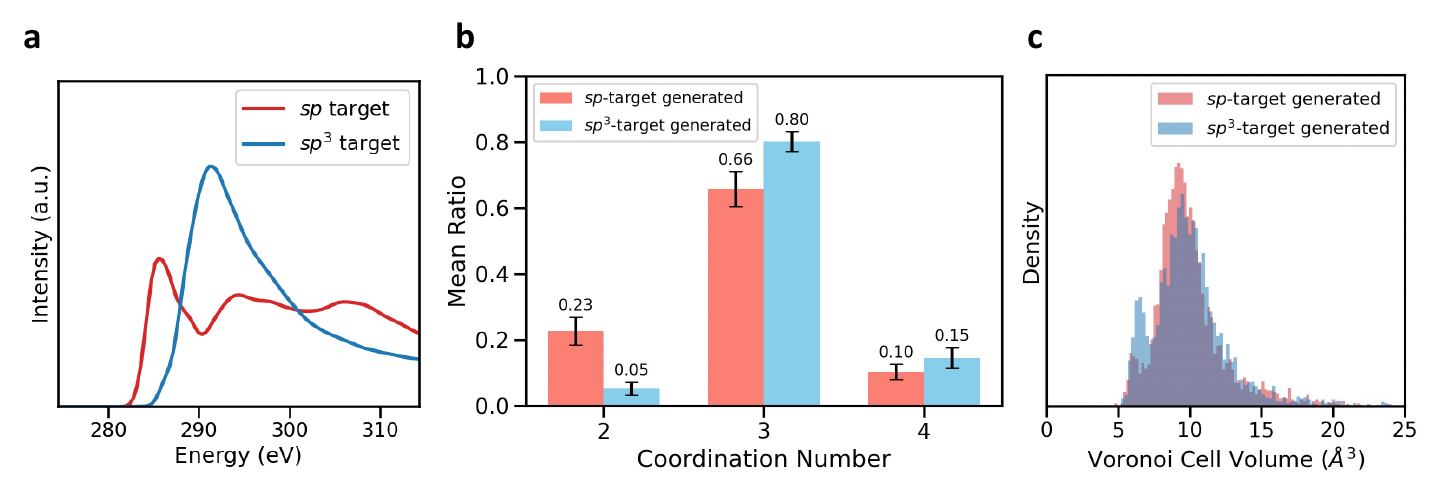}
    \caption{
    Demonstration of guided generation at the density of 2.0 g/cm$^3$ based on two distinctively different spectral guides/targets. As shown in the figure, a difference in (a) target XANES spectra certainly impacts the (b) coordination numbers and (c) Voronoi cell volume distribution of the generated $a$-C structures. Note that for each spectral input, 16 structure generations were performed, totaling 3,456 atoms. The standard deviation (across different generations) in the coordination number ratio is denoted by the error bar in (b). We also emphasize that $sp^2$ carbons (which have a coordination number of three) are still prevalent in both scenarios due to the strong conditioning effect of the density (2.0 g/cm$^3$), at which $sp^2$ hybridization is the typical topology.
    }
    \label{fig:cond-gen-ex1}
\end{figure}

We now validate the guided generation result of our conditional diffusion model. As previously stated, even without a given spectroscopy target, the $a$-C generation outcome depends on the user-specified density, to the extent that $sp$/$sp^2$ carbons dominate at low densities, and $sp^2$/$sp^3$ carbons dominate at high densities. In other words, density implicitly imposes a strong conditioning effect on the structural generation. Therefore, to investigate the conditioning effect of the spectroscopy guide alone, we provided two distinctively different spectral inputs while fixing the density (2.0 g/cm$^{3}$). One of the spectral guides is an averaged spectral curve from all $sp$ carbons, and the other from all $sp^3$ carbons in the training set. Effectively, these spectral inputs serve to guide the generation towards $sp$ or $sp^3$ topologies in structures that would otherwise be dominated by $sp^2$ carbons at 2.0 g/cm$^{3}$. As shown in Fig.~\ref{fig:cond-gen-ex1}, the difference in the two spectral inputs/targets certainly impacts the generation outcomes. Namely, generation guided by the $sp$ target expectedly results in a higher portion of two-fold coordination carbons, while the $sp^3$ target raises the portion of those with a four-fold coordination. Further, the $sp$-guided generation resulted in a stronger signature in the Voronoi cell volume distribution for the $sp$ topology \cite{jana2019structural}, while the $sp^3$-guided generation resulted in a stronger signature for the $sp^3$ topology. Overall, this demonstration suggests that we can effectively steer the generative process towards structures that are better aligned with the provided spectral input. For those interested in the precise matching of coordination number ratios, we repeated this procedure using density 1.5~g/cm$^3$, as illustrated in Supplemental Fig.~\ref{fig:density_1.5_coordination_number}.

It is crucial to point out that by design a XANES target might not be always fully achievable in the context of our conditional generative model. In this regard, it is more appropriate to consider XANES spectra as a guide for our diffusion model, where the generative process is biased towards certain topologies based on a given spectral target. As evident in Fig.~\ref{fig:cond-gen-ex1}b, the majority of the generated carbons remain $sp^2$ despite the ``target'' for $sp$ and $sp^3$ motifs. This is not at all an indication of failure for the generative model. In fact, at 2.0 g/cm$^{3}$, the typical topology is indeed $sp^2$, which the model observed during training. Therefore the model also imposes the prior knowledge about what should be realistically expected at 2.0 g/cm$^{3}$. Mathematically the generative process involves a balance between two terms: the unconditional term for achieving realistic samples and the guide term for biasing towards a target (further details in Methods). If all the generated carbons have $sp$ or $sp^3$ topology at 2.0 g/cm$^{3}$, then the resulting structures are very likely unphysical (such structures are also never observed in the MD training dataset).    

\begin{figure}
    \centering
    \includegraphics[width=0.7\textwidth]{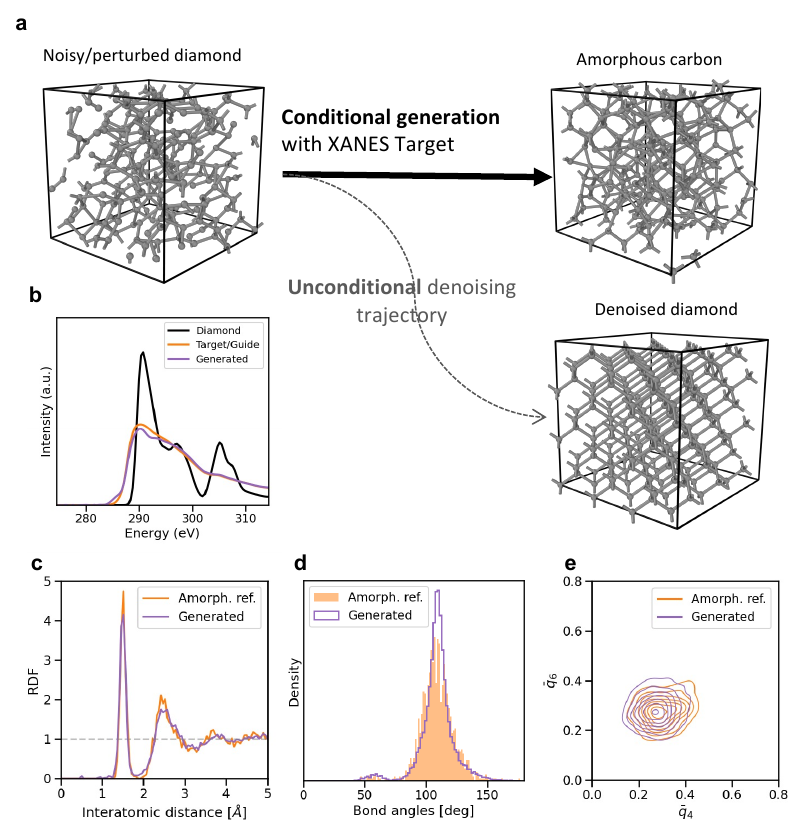}
    \caption{
    Demonstration of spectroscopy-guided generation where a heavily perturbed diamond structure can result in either perfect denoised diamond topology by default (unconditionally), or amorphous topology with the use of a spectral guide. (a) Visualization of the difference in the generation outcome with and without the spectral guide. (b) Comparison between the given target spectrum and the spectrum predicted from the generated amorphous structure, along with a diamond spectrum for reference. The generated amorphous structures are geometrically realistic compared to an MD-simulated reference structure (``amorph. ref.'') in terms of (c) RDF, (d) bond angle distribution, and (e) Steinhardt parameters.
    }
    \label{fig:cond-gen-ex2}
\end{figure}

As another demonstration of guided generation, we initialized a heavily perturbed diamond structure that would by default be denoised into perfect diamond via the unconditional denoising trajectory. We note that a separate generative model was trained in this case, where the training dataset includes only the perfect diamond structure and an amorphous structure of the same density (3.5 g/cm$^{3}$). As shown in Fig.~\ref{fig:cond-gen-ex2}a-b, by introducing a spectral target for the amorphous structure, the guided generation resulted in amorphous topology instead of the default diamond. Further, the generated amorphous structures are realistic in terms of RDF, bond angle distributions, and Steinhardt parameters (Fig.~\ref{fig:cond-gen-ex2}c-d). The choice of demonstrating the contrast between $a$-C and diamond is to further validate the conditioning effect of spectroscopy-guided generation, particularly for generating materials that may be considered completely different classes, e.g., amorphous vs. crystalline. However, generating crystalline motifs (at arbitrary scales) is likely an entirely different application scope. While our diffusion model can denoise heavily perturbed diamond structure into a virtually noiseless state, we did not thoroughly investigate the feasibility of denoising fully random structures into crystalline materials. We leave such a task for future work. The perturbed diamond structures were initialized by applying Gaussian noise $\epsilon \sim \mathcal{N}(0, 0.4^2)$ to the atomic coordinates (unit in \AA) of diamond structures. Lastly, as a bonus example, we also demonstrate the guided generation for $a$-C vs. buckyballs (Supplemental Fig.~\ref{fig:buckyball}). Moreover, through conditional generation, we observed the generation of structures with a diamond-like appearance from heavily noisy diamond input. The guide provided by conditional generation allows for further refinement of this structure, closely aligning it with the diamond structure (Supplemental Fig.~\ref{fig:refining}).

\begin{figure}
    \centering
    \includegraphics[width=0.7\textwidth]{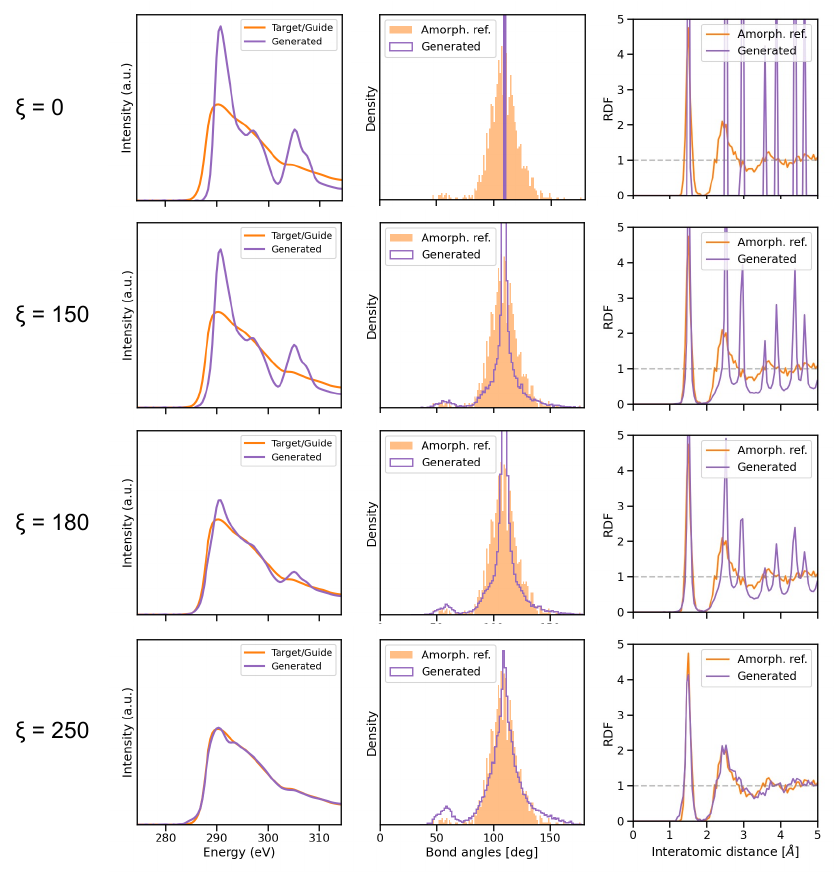}
    \caption{
    Based on the experiment for generating $a$-C vs. diamond, the tuning parameter $\xi$ can significantly impact the extent to which the generated structure is amorphous. Each row corresponds to a different value of $\xi$. The XANES spectrum (which is used as the target), the bond angle distribution, and the RDF from a reference amorphous structure are fixed in all rows. As $\xi$ increases, the generation transitions from the unconditional regime, where diamond is generated by default, to the conditional regime, where $a$-Cs are generated. 
    }
    \label{fig:xi-tuning}
\end{figure}

Finally, we briefly analyze the extent to which guided generation adheres to a given spectral target. Mathematically the conditional generative process is based on the conditional score function, which is the summation of two terms: the unconditional score points towards realistic samples, and the likelihood score points towards samples that satisfy the given target. The relative weights of the two terms can be adjusted with a tuning parameter $\xi$, as shown in Equation~\ref{eq:cond-score} and detailed in Section~\ref{sec:cond-diff}. A larger $\xi$ translates to more weight for the likelihood term, or a more strict emphasis for the target to be satisfied, while $\xi = 0$ signifies unconditional generation. One practical application of tuning the value of $\xi$ is to relax the target constraint such that more diverse samples can be generated. Based on the same experiment for generating $a$-C vs. diamond, the value of $\xi$ is shown to clearly impact the extent to which the generated structure is amorphous, as illustrated in detailed in Fig.~\ref{fig:xi-tuning}. As $\xi$ increases from zero, the generation transitions from the unconditional regime, where diamond is generated by default, to the conditional regime, where $a$-Cs are generated. Notably, the predicted spectrum from the generated carbons at $\xi =$ 180--250 closely matches the given target spectrum. This analysis is another validating evidence of conditional generation, particularly as a tunable knob for flexible inverse design of materials.

\section{Conclusions}

We propose a conditional generative framework based on a diffusion model to predict 3D atomic structures from spectroscopic data. This framework can be generalized to different targets that fit to a broader field of materials discovery by inverse design. First, we show that the generative model exhibits a remarkable scale-agnostic property, enabling it to learn from small-scale MD trajectories and generate large-scale structures. This versatility is useful for scenarios where accurate prediction of material properties is contingent on the generation of sufficiently large domains with diverse topologies. Second, our conditional generation process, influenced by spectroscopy input, offers control over the composition and structure of generated molecules. Therefore, one can guide the generative process to produce specific atomic arrangements, thereby tailoring materials for desired properties and applications. Lastly, through extensive fidelity assessments, we demonstrated the fidelity of our generative model. Specifically, the prior model accurately replicates key structural features observed in training data, while our posterior model yields structures that closely align with target data, including spectral predictions and structural characteristics.

Our work represents a significant stride in bridging the gap between materials characterization and atomic structure determination. In particular, by leveraging advanced generative models, our approach offer a promising solution to address to longstanding challenges in reverse engineering atomic structures from complex spectroscopic data. Our methodology holds promise for a multitude of future endeavors. For example, beyond the current scope, it can be applied to multiple  targets, including X-ray Photoelectron Spectroscopy, pair distribution function, and others, opening new avenues for predicting atomic structures from multi-modal materials characterizations. Finally, our framework can be leveraged for materials discovery, not only in the realm of spectroscopy but also in exploring various material properties as targeted. Examples include mechanical strengths of materials under extreme conditions, electrocatalytic activity of catalysts for hydrogen production, and permeability of ion-selective membranes, and so on.  
\section{Methods}

The methodologies employed in this work are described below, including disordered/amorphous structure generation via unconditional diffusion model, its extension to conditional/guided diffusion model formulation with the use of a forward model, how the models are trained, their GNN architectures, and the spectroscopy simulations performed for gathering data.

\subsection{Unconditional diffusion model}

We approach the task of amorphous/disordered structure generation with a machine learned generative model, namely the diffusion model \added{ \cite{Sohl-Dickstein2015-DPM, Ho2020-DDPM, Song2020-unified}}. In what may be considered the standard formulation, the diffusion model establishes a mapping between ``clean'', original data distribution $p_0(\mathbf{x}_0)$ and a latent distribution $p_t(\mathbf{x}_t)$ by iteratively applying (Gaussian) noises to $\mathbf{x}_0$, over a fictitious time quantity $t$, towards what is typically a unit Gaussian distribution $\mathcal{N}(0, \mathbf{I})$. This \textit{forward noising} process can be written as $p_t(\mathbf{x}_t | \mathbf{x}_0) = \mathcal{N}(\alpha_t \mathbf{x}_0, \sigma^2_t \mathbf{I})$ and implemented as $\mathbf{x}_t =\alpha_t\mathbf{x}_0  + \sigma_t \epsilon, ~ \epsilon \sim \mathcal{N}(0,\mathbf{I})$, where $(\alpha_t, \sigma_t)$ are noise schedule functions that dictate how noise is applied with respect to $t$. Typically, the span of $t$ is set to $(0, 1)$, over which $\alpha_t$ monotonically decreases from 1 to 0, and $\sigma_t$ monotonically increases from 0 to 1. The latent distribution $p_t(\mathbf{x}_t)$ therefore becomes a unit isotropic Gaussian at $t = 1$. The same forward noising process can also be described as Langevin dynamics with the following stochastic differential equation (SDE):
\begin{gather}
    \text{d}\mathbf{x}_t = f(t) \mathbf{x}_t \text{d}t + g(t) \text{d}\mathbf{w}_t, \\
    f(t) = \frac{\text{d} \log \alpha_t} {\text{d} t}, g^2(t) = \frac{\text{d} \sigma_t^2}{\text{d} t} - 2\frac{\text{d}\log \alpha_t}{\text{d} t} \sigma_t^2,
\end{gather}
where $f(t)$ and $g(t)$ are drift and diffusion coefficients related to $(\alpha_t, \sigma_t)$, and $\mathbf{w}_t$ is the standard Wiener process.

To generate a realistic data point, one first samples from the unit Gaussian latent distribution $\mathbf{x}_t \sim \mathcal{N}(0, \mathbf{I})$ at $t = 1$ and then maps it back to the original data space at $t = 0$ via the \textit{reverse denoising} process. While the forward noising operation is trivial, the reverse denoising operation is not and requires a learned model. It is described as the following SDE:
\begin{align}
    \text{d}\mathbf{x}_t 
        &= \left[ f(t)\mathbf{x}_t - g^2(t) \nabla_{\mathbf{x}_t} \log p_t(\mathbf{x}_t) \right] \text{d}t + g(t) \text{d}\mathbf{w}_t, \\
        &\approx \left[ f(t)\mathbf{x}_t - g^2(t) s_{\theta}(\mathbf{x}_t, t) \right] \text{d}t + g(t) \text{d}\mathbf{w}_t, \\
        &=       \left[ f(t)\mathbf{x}_t + \frac{g^2(t)}{\sigma_t} \epsilon_{\theta}(\mathbf{x}_t, t) \right] \text{d}t + g(t) \text{d}\mathbf{w}_t,
\end{align}
where $\nabla_{\mathbf{x}_t} \log p_t(\mathbf{x}_t)$ is the so-called score function and in this work is approximated by the score model $s_\theta(\mathbf{x}_t, t)$ with parameters $\theta$. The score function can also be approximated using $\epsilon_{\theta}$, which learns to predict the noise $\epsilon$ in noisy $\mathbf{x}_t$, and is related to $s_{\theta}$ by $\epsilon_{\theta} = - \sigma_t s_{\theta}$. The reverse denoising SDE can then be solved using well known techniques such as the Euler-Maruyama method \cite{mao2015truncated}.

While the above formulation of the diffusion model is generally applicable to many data types, it is modified in this work to account for atomic structures in periodic cell boundaries. Note that an atomic structure may be generally represented as a tuple of the atomic coordinates and atom types ($\mathbf{r}$, $\mathbf{z}$), along with unit cell dimensions $\mathbf{A} = (\mathbf{a}_1, \mathbf{a}_2, \mathbf{a}_3) \in (\mathbb{R}^3, \mathbb{R}^3, \mathbb{R}^3)$ if the structure is periodic. However, the amorphous carbon systems studied in this work are single-element only (carbon) and therefore $\mathbf{z}$ is omitted, leaving $\mathbf{x} = \mathbf{r}$. Since $\mathbf{r}$ lies in a unit cell with periodic boundaries, it is sensible to apply the ``variance-exploding'' scheme for the noise schedule, where $\alpha_t$ is fixed to 1 and $\sigma_t$ monotonically increases from 0 to some maximum value $k$. This means that the drift coefficient becomes zero $f(t) = 0$. The updated forward noising SDE is then
\begin{equation}
    \text{d}\mathbf{x}_t = g(t) \text{d}\mathbf{w}_t, ~ g^2(t) = \frac{\text{d} \sigma_t^2}{\text{d} t},
\end{equation}
and its corresponding reverse denoising SDE is
\begin{equation}
    \text{d}\mathbf{x}_t = - g^2(t) \nabla_{\mathbf{x}_t} \log p_t(\mathbf{x}_t) \text{d}t + g(t) \text{d}\mathbf{w}_t.
\end{equation}
Importantly, due to the periodic boundaries, the noisy, perturbed structure becomes asymptotically a uniformly random structure. Therefore, to sample a realistic structure, one instead initializes a set of uniformly random coordinates $\mathbf{r}$ inside a periodic unit cell---rather than Gaussian-distributed $\mathbf{r}$---and then applies the reverse denoising SDE. The structure generation algorithm is listed in Algorithm \ref{alg:denoise-SDE}. In this work we define the noise schedule as $(\alpha_t, \sigma_t) = (1, kt)$, where the span of $t$ is still set to $(0, 1)$. Notably, the structure generation algorithm is \textit{scale-agnostic}, allowing for users to specify the unit cell dimensions $\mathbf{A}$ and structure density $D$, which influence the number of atoms $N_a$ for the generation. In practice, $D$ should have a realistic value, similar to the densities of the structures in the training dataset.

\begin{algorithm}
\caption{Unconditional disordered structure generation via Euler-Maruyama method} \label{alg:denoise-SDE}
\begin{algorithmic}
    \Require{
        Periodic unit cell $\mathbf{A} = (\mathbf{a}_1, \mathbf{a}_2, \mathbf{a}_3)$,
        density $D$,
        learned score model $s_{\theta}$,
        noise schedule $g(t)$,
        and timestep size $\Delta t$ (which has a negative value).
    }
    \State Determine number of atoms $N_a$ from $\mathbf{A}$ and $\rho$
    \State $\mathbf{x} \gets \mathcal{U}(\Omega)$ for $N_a$ atoms, where $\Omega$ is the domain inside $\mathbf{A}$
    \State $t \gets 0.999$
    \Repeat
        \State $\Delta \mathbf{x} \gets -g^2(t) s_{\theta}(\mathbf{x}, t) \Delta t + g(t) \epsilon, ~ \epsilon \sim \mathcal{N}(0, |\Delta t|\mathbf{I})$
        \State $\mathbf{x} \gets \mathbf{x} + \Delta \mathbf{x}$
        \State $t \gets t + \Delta t$
    \Until{$t \approx 0$}
\end{algorithmic}
\end{algorithm}

\subsection{Conditional diffusion model} \label{sec:cond-diff}

The theory of the diffusion model, based on an underlying score function, can be quite naturally extended for conditional generation. First, the score function introduced in the previous section is written as $\nabla_{\mathbf{x}_t} \log p_t(\mathbf{x}_t)$ and is regarded as the unconditional, unguided, or prior score function, i.e., without conditioning bias or guide during the generation/denoising process. Then, consider a conditioning or posterior score function $\nabla_{\mathbf{x}_t} \log p_t(\mathbf{x}_t | y)$ that is conditioned on some target or guide property $y$. Importantly, this posterior score can be decomposed via Bayes' rule into
\begin{align}
    \nabla_{\mathbf{x}_t} \log p_t(\mathbf{x}_t | y) 
        & = \nabla_{\mathbf{x}_t} \log p_t(\mathbf{x}_t) + \nabla_{\mathbf{x}_t} \log p_t(y | \mathbf{x}_t) \\
        & \approx  s_{\theta}(\mathbf{x}, t) + \nabla_{\mathbf{x}_t} \log p_t(y | \mathbf{x}_t).
\end{align}

It is now apparent that conditional generation enabled by the diffusion model can be very flexible, since computing the posterior score amounts to simple addition of the decoupled prior score $\nabla_{\mathbf{x}_t} \log p_t(\mathbf{x}_t)$---approximated by $s_{\theta}$---and the likelihood score $\nabla_{\mathbf{x}_t} \log p_t(y | \mathbf{x}_t)$. Further, computing the likelihood score entails taking the gradient of (the log of) a probabilistic forward prediction model $p_t(y | \mathbf{x}_t)$. For different conditional quantities $y$ such as spectral signatures, Steinhardt parameters, band gap, and so on, different forward models can be combined with the prior model in a plug-and-play fashion. In the case of a deterministic forward model $y = F(\mathbf{x})$, converting it into the probabilistic counterpart can be straightforward by adding Gaussian noises of variance $\hat{\sigma}^2$ to the forward prediction: $y = F(\mathbf{x}) + \epsilon, ~ \epsilon \sim \mathcal{N}(0, \hat{\sigma}^2 \mathbf{I})$. The resulting likelihood score is then
\begin{equation}
    \nabla_{\mathbf{x}_t} \log p_t(y | \mathbf{x}_t) = \\
        - \frac{1}{2\hat{\sigma}^2} \nabla_{\mathbf{x}_t} \| y - F(\mathbf{x}_t) \|^2,
\end{equation}
which can be interpreted as pointing to lower values of the L2 loss between the forward prediction $ F(\mathbf{x}_t)$ and a given target $y$. In terms of implementation, the conditional generation process is virtually identical to Algorithm~\ref{alg:denoise-SDE} except for the input score model, which is replaced by $\nabla_{\mathbf{x}_t} \log p_t(\mathbf{x}_t | y)$.

However, there is still a caveat with the conditional generation formulation: evaluating the forward prediction $F(\mathbf{x}_t)$ involves noisy input $\mathbf{x}_t$ but the forward model may be ill-defined with respect to noisy, highly perturbed $\mathbf{x}_t$, which at $t = 1$ is complete noise without any meaningful signal. For some forward models, this does not pose as an issue, such as the case of atomic descriptors/fingerprints, e.g., Steinhardt\cite{Steinhardt1983PRB}, ACSF\cite{behler2011atom}, SOAP\cite{bartok2013representing}, MBTR\cite{huo2017unified} order parameters, where the structure-to-descriptor forward mapping still has a valid output when $\mathbf{x}_t$ is highly noisy or even completely random. The forward model developed in this work, however, is a graph network model trained to predict spectroscopic curves off of atomic motifs in disordered or amorphous carbons, which still exhibit certain topological order and are far from completely random. Our forward model output is therefore ill-defined and likely holds no meaningful gradient information with noisy input. To address this, Chung et al.\cite{chung2022diffusion} proposed to inverse the relationship $\mathbf{x}_t = \alpha_t \mathbf{x}_0 + \sigma_t \epsilon$ and estimate the ``clean'' version of data $\hat{\mathbf{x}}_0$ from noisy $\mathbf{x}_t$ as
\begin{align}
    \hat{\mathbf{x}}_0 
        &= \frac{1}{\alpha_t} \left( \mathbf{x}_t - \sigma_t \epsilon_{\theta}(\mathbf{x}_t, t) \right) \\
        &= \frac{1}{\alpha_t} \left( \mathbf{x}_t + \sigma^2_t s_{\theta}(\mathbf{x}_t, t) \right),
\end{align}
followed by forward model prediction, resulting in the following likelihood score:
\begin{equation} \label{eq:cond-score}
    \nabla_{\mathbf{x}_t} \log p_t(y | \mathbf{x}_t) \\
        = - \rho \nabla_{\mathbf{x}_t} \| y - F(\hat{\mathbf{x}}_0(\mathbf{x}_t)) \|^2 
        = - \frac{\xi} {\| y - F(\cdot) \|} \nabla_{\mathbf{x}_t} \| y - F(\hat{\mathbf{x}}_0(\mathbf{x}_t)) \|^2,
\end{equation}
where the stepsize $\rho$ is another implementation practice by Chung et al.\cite{chung2022diffusion}, and is defined as a normalized quantity $\rho = \xi / \| y - F(\hat{\mathbf{x}}_0(\mathbf{x}_t)) \|$, with $\xi$ being a tuning parameter. In this work, we have found the conditional generation with the spectroscopy forward model to be viable only with this ``estimation trick''.

\subsection{Forward model}

As previously stated, the forward model used in this work for the purpose of conditional or guided generation of disordered carbons is a spectroscopy prediction model. Specifically, it is a GNN that outputs per-atom spectroscopic curves based on local atomic motifs. In this representation, the atomic structure is represented as a graph, wherein nodes encapsulate information pertaining to atomic species and atomic positional coordinates, while edges encode atomic bonding information in the form of vectorized interatomic distances. The GNN is subjected to training on a dataset containing atomistic structures along with their corresponding XANES spectra.

\subsection{Model training}

The score model $s_{\theta}(\mathbf{x}_t, t)$ is a GNN trained with dataset $\mathcal{D}_s = \{(\mathbf{x}_0, \mathbf{A})^{(i)}\}$ that consists of MD snapshots containing atomic coordinates $\mathbf{x}_0$ and unit cell dimensions $\mathbf{A}$, with atom type information $\mathbf{z}$ omitted due to the dataset being single-element. The score is learned via the denoising score matching loss
\begin{equation}
    \begin{aligned}
        \mathbf{x}_t 
            &= \alpha_t \mathbf{x}_0 + \sigma_t \epsilon, ~ \epsilon \sim \mathcal{N}(0,\mathbf{I}), \\
        \mathcal{L}_\text{DSM} 
            &= \mathbb{E}_{\mathbf{x}_0, \epsilon, t} \left[ 
                \Vert s_{\theta}(\mathbf{x}_t , t) \sigma_t + \epsilon \Vert^2
            \right],
    \end{aligned}
\end{equation}
where $\mathbf{A}$ is implicitly accounted for by the atomic graph representation. Algorithm~\ref{alg:train-score} lists the pseudocode for training the score model.

\begin{algorithm}
\caption{Score model training} \label{alg:train-score}
\begin{algorithmic}    
    \Require{Training dataset $\mathcal{D}_s$,
        noise schedule $(\alpha_t, \sigma_t)$,
        initial model parameters $\theta$,
        another set of model parameters $\theta'$ for EMA,
        EMA decay rate $\beta$,
        gradient descent optimizer \verb|Optim|,
        and learning rate $\eta$.
    }
    \Repeat
        \State Sample $(\mathbf{x}_0, \mathbf{A}) \sim \mathcal{D}$
        \State Sample $t \sim \mathcal{U}(0.001, 0.999)$
        \State Sample $\epsilon \sim \mathcal{N}(0, \mathbf{I})$
        \State $\mathbf{x}_t \gets \alpha_t \mathbf{x}_0 + \sigma_t \epsilon, ~ \epsilon \sim \mathcal{N}(0,\mathbf{I})$
        \State $\mathcal{L}_\text{DSM} 
            \gets \mathbb{E}_{\mathbf{x}_0, \epsilon, t} \left[ 
                \Vert s_{\theta}(\mathbf{x}_t , t) \sigma_t + \epsilon \Vert^2
            \right]$
       \State $\theta \gets \verb|Optim|(\mathcal{L}_{\text{DSM}}, \theta, \eta)$
       \State $\theta' \gets \beta \theta' + (1-\beta)\theta $
    \Until{convergence}
\end{algorithmic}
\end{algorithm}

The forward model is also a GNN, trained with dataset $\mathcal{D}_F = \{(\mathbf{x}_0, \mathbf{A}, y)^{(i)}\}$ that additionally has per-atom spectroscopic curves $y$. The spectroscopic prediction is learned via a simple mean-squared-error (MSE) loss between the predicted output $\hat{y}$ and $y$: $\mathcal{L}_\text{MSE} = \text{MSE}(\hat{y}, y)$.

For both the score model and the forward model, exponential moving average (EMA) was used to faciliate training. Rectified Adam (RAdam) was used as the gradient descent optimizer without learning rate or weight decay. For every sampled minibatch of the structure snapshots, the atomic coordinates are randomly rotated in order for the score, (forward) model to learn rotational equivariance (invariance).
The training parameters are detailed in Table \ref{tab:train-params}. All other parameters, if unspecified, default to implementations in PyTorch 1.11.0~\cite{paszke2019pytorch} and PyTorch-Geometric 2.0.4~\cite{fey2019fast}, which were used for the GNN implementations and training.

\subsection{GNN architecture}

Both the score model and the forward model share the same GNN architecture design, which consists of three components: an Encoder, a Processor (for graph message-passing), and a Decoder. The input atomic graph to these operations is denoted as $(V, E)$, where $V$ is a set node features $\{\mathbf{h}_i\}$ for nodeatom $i$, and $E$ is a set of edge/bond features $\{\mathbf{e}_{ij}\}$ from node $i$ to node $j$. The set of edges $E$ is determined by a cutoff distance $r_c$, within which a pair of nodes form an edge connection.

In the Encoder stage, the node features are computed as a combined embedding of the atom type $\mathbf{z}$ and the fictitious time $t$:
\begin{equation}
    \mathbf{h}_i \gets M_H(\text{onehot}(\mathbf{z}_i)) + M_T(\text{GaussianRFF}(t)),
\end{equation}
where $M_H$ and $M_T$ are multilayer perceptrons (MLPs) that embed the one-hot representation of $\mathbf{z}$ and the Gaussian Fourtier features \cite{tancik2020fourier} (GaussianRFF) of $t$, respectively. Note that in this work, $\mathbf{z}$ is simply a dummy variable because the atom type information in the amorpohus carbon dataset is omitted, but the Encoder is defined as above so that it can be adjusted to account for different atom types. The edge feature $\mathbf{e}_{ij}$ is the concatenation of the edge vectors $\mathbf{r}_{ij} = \mathbf{r}_j - \mathbf{r}_i$ and the corresponding edge lengths $\bar{r}_{ij}$, followed by an MLP $M_E$:
\begin{equation}
    \mathbf{e}_{ij} \gets M_E(\mathbf{r}_{ij} \oplus \bar{r}_{ij}),
\end{equation}
where $\oplus$ designates concatenation of features. $M_H$, $M_E$, and $M_T$ consist of two dense linear layers, with SiLU activation function after the first layer and layer normalization after the second.

We used the Processor from MeshGraphNets \cite{Pfaff2020} for implementing graph convolutions. We refer to the original paper for further details of MeshGraphNets. There is a main difference between our implementation and the original work: the original work considers two sets of edges (mesh-space edges and world-space edges), whereas there is only one in our work.

The Decoder is a simple MLP $M_O$ that maps the node features into prediction output:
\begin{equation}
    \hat{y}_i \gets M_O(\mathbf{h}_i).
\end{equation}
For the score model, the output is the approximated score function, and for the forward model, the output is the spectroscopic signature. $M_O$ has the same architecture as that of $M_H$, $M_E$, and $M_T$, except that there is no layer normalization at the end.

Model parameters are listed in Table \ref{tab:train-params}.

\begin{table}
    \centering
    \caption{Training and model parameters.}
    \begin{tabular}{l l l}
    \toprule
    Name & \multicolumn{2}{c}{Value} \\
         & Score model & Forward model \\
    \midrule
    Number of model updates             & 1,000,000  &  400,000 \\
    Learning rate $\eta$                & 0.001       & 0.0001  \\
    EMA decay rate $\beta$              & 0.999      &  0.999 \\
    Minibatch size                      & 64         &  8 \\
    Node, edge, time feature dimensions & 256        &  256 \\
    Number of graph convolutions        & 5          &  3 \\
    \bottomrule
    \end{tabular}
    \label{tab:train-params}
\end{table}
\subsection{Restart sampling}
\begin{figure}[t]
    \centering
    \includegraphics[width=0.5\textwidth]{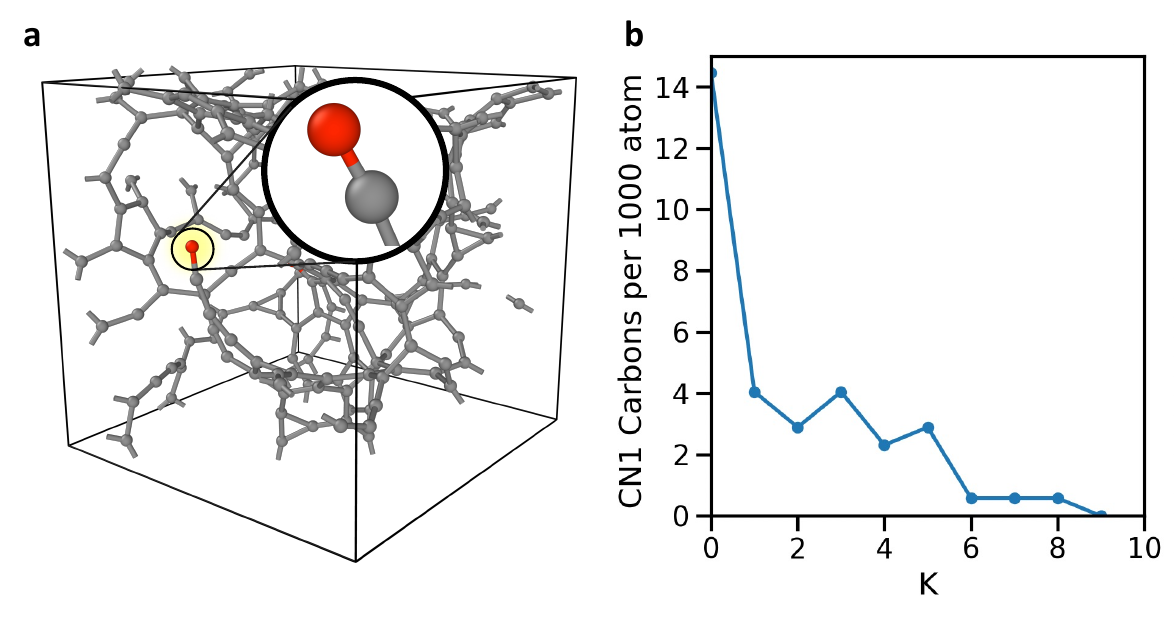}
    \caption{Unphyiscal CN1 (coordination number of 1) carbons can be effectively reduced by $K$ cycles of restart sampling. An example of CN1 carbon is shown in (a). Roughly 6--10 cycles of restart sampling can effectively minimiize occurances of CN1 carbons (b).
    }
    \label{fig:restart-sample}
\end{figure}

Since the generative model in this work is not constrained by physics rule, there is always a chance for the model to generate unphysical topologies that are extremely unlikely or virtually impossible. Such ``bad'' samples occur since there are always errors associated with any fitted score model. In this work one main instance of unphysical $a$-C topology is carbons with a coordination number of only one (CN1), which can exist in our generation results using the usual SDE sampler. To address this, we have found the \textit{restart sampling} method \cite{xu2023restart} to be effective in avoiding CN1 carbon generation (Fig.~\ref{fig:restart-sample}). In brief, the restart sampling method entails multiple cycles of reapplying forward noising and reverse denoising operations between two points in the denoising time schedule. In this work, the two points are  $t = 0, 0.3$. Note that we apply SDE denoising, but the original restart sampling technique is based on ODE (ordinary differential equation) denoising.
\subsection{X-ray absorption spectral simulations}
We extracted a dataset consisting of 58 atomic configurations of $a$-C at different densities from molecular dynamics (MD) simulations with the Gaussian approximation potential (GAP), which has been shown to provide a good description of structural properties of $a$-C at the room temperature \cite{deringer2017machine}. Each configuration consists of 216 atoms in a simple-cubic lattice with varying densities of 1.5, 2.0, 2.5, 3.0, and 3.5 g/cm$^{3}$ to sample a wide range of local carbon structures. Collectively, this dataset encompassed a total of 12,528 local carbon sites. We then calculated K-edge spectra of these atomic sites using Density Functional Theory (DFT) with the core-hole approximation (XCH) method \cite{prendergast2006x}. All computational procedures were carried out utilizing the QuantumESPRESSO package \cite{giannozzi2009}. To efficiently sample the Brillouin zone, we employed the Shirley reduced basis method \cite{prendergast2009}. The Perdew–Burke–Ernzerhof (PBE) formulation of the generalized-gradient approximation (GGA) was adopted \cite{perdew1996}. We represented electronic wavefunctions and charge density using a plane-wave basis set with kinetic energy cutoffs of 30 Ry and 240 Ry, respectively. All our calculations utilized ultrasoft pseudopotentials \cite{vanderbilt1990}. For the excited atom, we employed a modified pseudopotential with one electron removed from the 1s core orbital.

Within the XCH approximation, we approximated the first core-excited state through constrained-occupancy DFT, which involved incorporating both the core-hole pseudopotential and the excited electron positioned at the conduction band minimum. Transition probabilities at each energy level were computed using Fermi's Golden Rule and then convoluted with a uniform Gaussian broadening of 0.2 eV. To align the calculated spectrum with experimental data \cite{sainio2020}, we matched the first major peak of the graphite spectrum and applied the same constant shift to align the spectra of all carbon sites within $a$-C systems. It is necessary to note that other methods can also be used to compute XANES spectra, such as those based on time-dependent DFT or more computationally intensive Bethe–Salpeter calculations, which explicitly consider the interaction between the electron and core-hole~\cite{shirley1998,vinson2012,lopata2012}.

\section*{Acknowledgements}
This work was carried out under the auspices of the US Department of Energy by Lawrence Livermore National Laboratory (LLNL) under contract No. DE-AC52-07NA27344. We acknowledge support from LLNL Laboratory Directed Research and Development (LDRD) Program Grant No. 22-ERD-014. This research used the Theory and Computation facility of the Center for Functional Nanomaterials, which is a US Department of Energy Office of Science User Facility at Brookhaven National Laboratory under Contract DE-SC0012704. M.R.C., and D.L. acknowledge the support by the U.S. Department of Energy, Office of Science, Office Basic Energy Sciences, under Award FWPPS-030. F.Z. was supported by LLNL LDRD Grant 23-SI-006. J.C acknowledges support from the College of Engineering and the Department of Mechanical Engineering at Boston University.

\section*{Author Contributions}
H.K., T.H., and T.A.P. supervised the research. F.A. carried out MD simulations of amorphous carbons. W.S. and W. J. computed the XANES spectra of amorphous carbons. F.Z. and T.H. developed the concept of conditional diffusion models. H.K., T.H., and T.A.P. wrote the manuscript with inputs from all authors.

\section*{Data Availability}
All data required to reproduce this work can be requested by contacting the corresponding authors.

\section*{Code Availability}
The source code for this work is available at https://github.com/LLNL/graphite.

\section*{Competing interests}
The authors state that there is no conflict of interest.

\printbibliography
\newpage
\setcounter{figure}{0}
\renewcommand{\figurename}{Supplementary Figure}
\renewcommand{\theHfigure}{Supplement.\thefigure}

\setcounter{table}{0}
\renewcommand{\tablename}{Supplementary Table}
\renewcommand{\theHtable}{Supplement.\thetable}


\begin{figure}
    \centering
    \includegraphics[width=\textwidth]{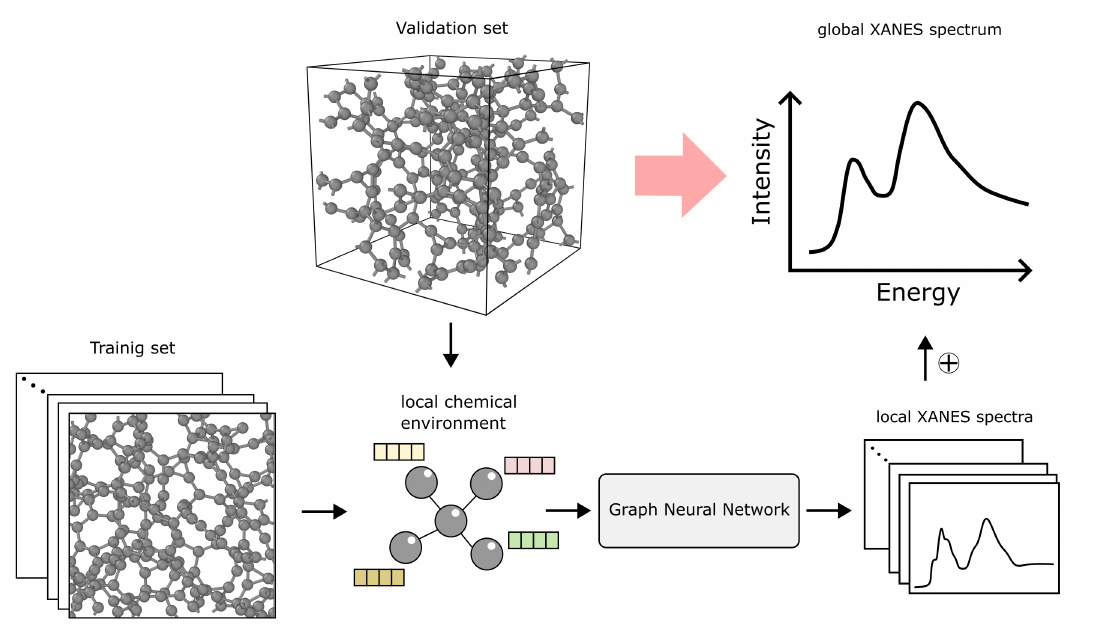}
    \caption{Schematic of forward model. This schematic elucidates the operation of our forward model, which receives local chemical environment data as input and employs an equivariant graph neural network for the production of local XANES spectra of individual atom. Subsequently, for validation purposes, we predict the local XANES spectra of all atoms in structures that were omitted from the training dataset. These locally predicted spectra are cumulatively summed to generate the global XANES spectra, allowing for a quantitative assessment of their concordance with the DFT computed global XANES spectrum.
    }
    \label{fig:forward_model_scheme}
\end{figure}

\begin{figure}
    \centering
    \includegraphics[width=0.8\textwidth]{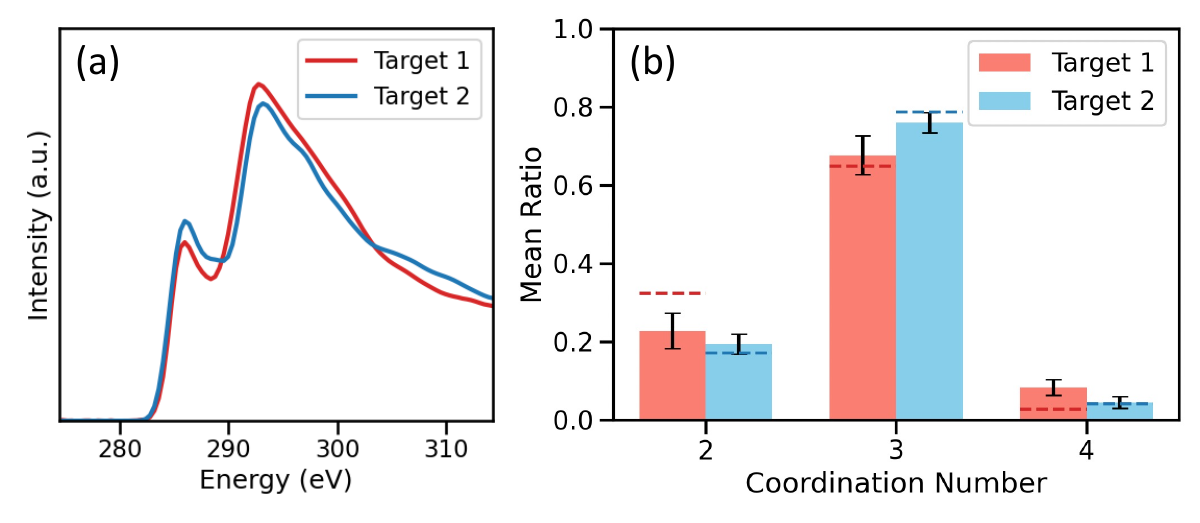}
    \caption{
    (a) Spectroscopy of two distinct structures with a density of 1.5~g/cm$^3$. Despite having the same density, their differing topologies result in distinct spectroscopic profiles. (b) Coordination number ratios of amorphous carbon structures generated through conditional generation based on the spectroscopies in panel (a). Dashed lines represent target coordination numbers, derived from real structures corresponding to the spectroscopies in panel (a). The coordination number ratios of the generated structures closely align with these target values, validating the fidelity of the conditional generation process.}
    \label{fig:density_1.5_coordination_number}
\end{figure}

\begin{figure}
    \centering
    \includegraphics[width=0.9\textwidth]{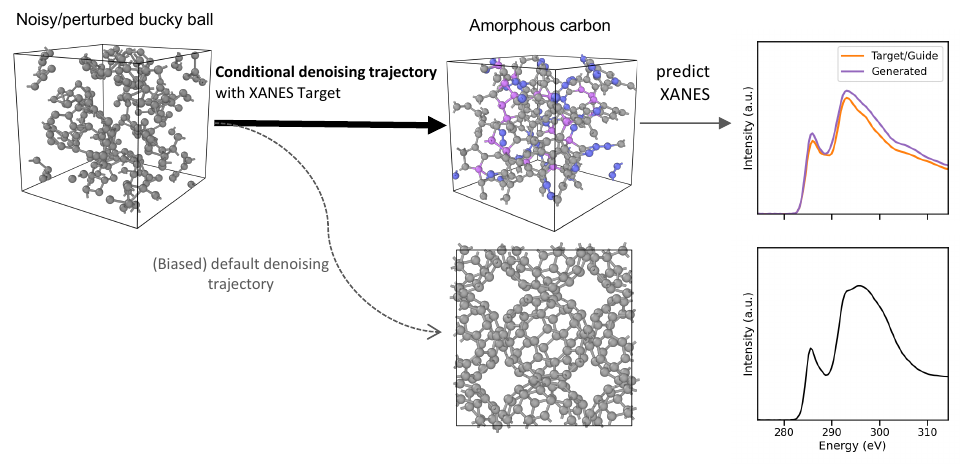}
    \caption{The generation process, starting from an initial noisy structure, unfolds in two distinct transformations: (1) conditional denoising trajectory, influenced by target XANES spectroscopy, and  (2) default denoising trajectory, driven by an unconditional diffusion model. Comparisons of the generated amorphous structures against target data, including spectroscopy predictions is shown on the right panel.
    }
    \label{fig:buckyball}
\end{figure}

\begin{figure}
    \centering
    \includegraphics[width=0.9\textwidth]{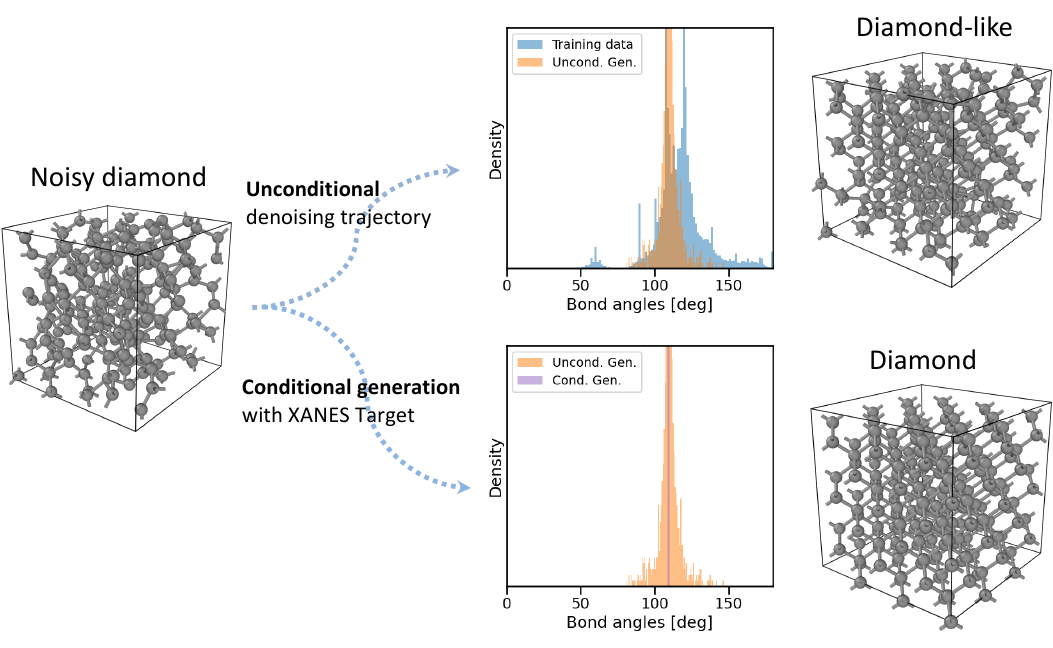}
    \caption{The top row shows the generation of diamond-like structure from initial heavily noisy diamond structure via unconditional generation, and the bottom row illustrates the refined structure achieved through conditional generation with the guide based on the diamond's spectrum. The conditional generation process significantly enhances the resemblance of the resulting structure to the ideal diamond structure. This showcases the utility of conditional generation in refining generated structures with specific spectroscopic targets.
    }
    \label{fig:refining}
\end{figure}

\end{document}